\documentclass[preprint,aps,pra,showpacs,floatfix]{revtex4}
\usepackage[dvips]{graphicx}

\usepackage{longtable}
\usepackage{dcolumn}
\usepackage[dvips]{graphicx}
\usepackage{bm}
\usepackage{bbm}

\usepackage{nicefrac}
\usepackage{amsmath}
\usepackage{amsfonts}
\usepackage{amssymb}
\usepackage{amsthm}

\newcolumntype{.}{D{x}{}{-1}}

\begin{document}

\newcommand{\vare}{\varepsilon}

\newcommand{\pr}{^{\prime}}
\newcommand{\bfx}{{\bf x}}
\newcommand{\bfy}{{\bf y}}
\newcommand{\bfz}{{\bf z}}
\newcommand{\bfp}{{\bf p}}
\newcommand{\la}{\langle}
\newcommand{\ra}{\rangle}
\newcommand{\eps}{\varepsilon}
\newcommand{\beq}{\begin{equation}}
\newcommand{\eeq }{\end{equation}}
\newcommand{\beqn}{\begin{eqnarray}}
\newcommand{\eeqn }{\end{eqnarray}}
\newcommand{\ba}{\begin{array}}
\newcommand{\ea}{\end{array}}
\newcommand{\balpha}{{\mbox{\boldmath$\alpha$}}}
\newcommand{\Za}{Z \alpha}
\newcommand{\aZ}{\alpha Z}
\newcommand{\etal}{{\it et al.}}

\newcommand{\lbr}{\langle}
\newcommand{\rbr}{\rangle}

\newcommand{\Dmatrix}[4]{
        \left(
        \begin{array}{cc}
        #1  & #2   \\
        #3  & #4   \\
        \end{array}
        \right)
        }

\title{Screened self-energy correction to the 2p$_{3/2}$-2s transition energy
in Li-like ions.}
\author{V.~A.~Yerokhin,$^{1,2}$
  A.~N.~Artemyev,$^{2,3}$ V.~M.~Shabaev,$^{2,3}$
  G. Plunien,$^{3}$  and G.Soff$^{3}$}
\affiliation{
$^1$ Center for Advanced Studies, St. Petersburg State Polytechnical
University, Polytekhnicheskaya 29, St. Petersburg 195251, Russia\\
$^2$Department of Physics, St. Petersburg State University, Oulianovskaya 1,
Petrodvorets, St. Petersburg 198504, Russia\\
$^3$Institut f\"{u}r Theoretische Physik, TU Dresden, Mommsenstra{\ss}e 13,
D-01062 Dresden, Germany
}

\begin{abstract}

We present an {\it ab initio} calculation of the screened self-energy
correction for $(1s)^2 2p_{3/2}$ and $(1s)^2 2s$ states of Li-like ions with
nuclear charge numbers in the range $Z = 12$-100. The evaluation is carried
out to all orders in the nuclear-strength parameter $\Za$. This investigation
concludes our calculations of all two-electron QED corrections for the
$2p_{3/2}$-$2s$ transition energy in Li-like ions and thus considerably
improves theoretical predictions for this transition for high-$Z$ ions.

\end{abstract}
\pacs{12.20.Ds, 31.30.Jv, 31.10.+z}

\maketitle

\section*{Introduction}

Recent progress in high-precision spectroscopy of highly charged ions has
attracted significant attention to these systems. Accurate experimental
determination of energy levels is nowadays possible for very heavy ions of
the periodical table up to H-like uranium. Such systems provide a unique
testing ground for quantum electrodynamics (QED) in a strong field of the
nucleus. While the simplicity of the hydrogen isoelectronic sequence makes it
ideal for theoretical investigations, the achieved experimental accuracy is
by far better in few-electron systems. In particular, the Lamb shift in
H-like uranium is presently known at the 3\% level \cite{stoehlker:00:prl},
whereas the $2p_{3/2}$-$2s$ splitting in Li-like bismuth was measured to be
\beq
E_{2p_{3/2}}-E_{2s} = 2788.139(9)~{\rm eV}\,,
\eeq
where the QED correction, which is about 26~eV, can be (in principle) tested
at the 0.15\% level. Very high experimental accuracy is also achieved for
$2p_{3/2}$-$2s$ and $2p_{1/2}$-$2s$ transitions in other heavy Li-like ions
\cite{schweppe:91,bosselmann:99,feili:00,brandau:04}.

In order to match the experimental accuracy for Li-like ions in theoretical
investigations, rigorous calculations of all QED effects to second order in
the fine structure constant $\alpha$ are needed. Characteristic property of
heavy ions is that the nuclear-strength parameter $\Za$ can not be used as an
expansion parameter in theoretical considerations, which corresponds to the
non-perturbative (in $\Za$) regime of the bound-state QED. On the other hand,
the electron-electron interaction in heavy ions can be accounted for by a
rapidly-converging perturbation expansion in the parameter $1/Z$. In our
approach, we start in zeroth approximation with non-interacting electrons
propagating in the external field of the nucleus (the Furry picture).
Corrections to this approximation arise from exchanges by one, two, and more
virtual photons. For corrections involving one and two virtual photons, we
employ a rigorous QED treatment complete to all orders in the parameter
$\Za$. Higher-order corrections can not be addressed within bound-state QED
at present. This part should be, therefore, evaluated within the standard
approach based on the no-pair Dirac-Breit Hamiltonian.

Such project was carried out for the $2p_{1/2}$-$2s$ transition of Li-like
ions in a series of our investigations
\cite{artemyev:99,yerokhin:99:sescr,zherebtsov:00,yerokhin:00:prl,yerokhin:01:2ph}.
Comparison of the total theoretical value obtained for the Li-like uranium
\cite{yerokhin:00:prl,yerokhin:01:2ph} with the corresponding experimental
results \cite{schweppe:91,brandau:04} probes QED effects of {\em second}
order in $\alpha$ on the level of about 17\%. This is the strictest test of
predictions of bound-state QED in the background of a strong external field
at the moment. Our present goal is to perform a similar project for the
$2p_{3/2}$-$2s$ transition energy in Li-like ions. For one specific ion,
Li-like bismuth, analogous calculation was carried out previously by
Sapirstein and Cheng \cite{sapirstein:01:lamb}.

The leading ($\sim 1/Z^0$) QED effect in Li-like ions is the one-loop
self-energy and vacuum-polarization corrections. Their calculation can be
considered as well established at present (see, e.g., review
\cite{mohr:98:rev} and references therein). Other important QED effects are
those of order $1/Z$, the {\em screened$\,$} self-energy and
vacuum-polarization corrections and the two-photon exchange correction. The
screened vacuum-polarization correction for $n=1$ and $n=2$ states of Li-like
ions was evaluated previously in Ref.~\cite{artemyev:99}. The two-photon
exchange correction was obtained in
Refs.~\cite{yerokhin:00:prl,yerokhin:01:2ph} for the $2s$ and $2p_{1/2}$
states and in Refs.~\cite{sysak:02,artemyev:03} for the $2p_{3/2}$ state. An
independent evaluation of the two-photon exchange correction for Li-like ions
was performed by Andreev \etal$\,$ \cite{andreev:01,andreev:03}. A
calculation of the screened self-energy correction was carried out in
Ref.~\cite{yerokhin:99:sescr} for the $2s$ and $2p_{1/2}$ states. The goal of
the present investigation is to perform an evaluation of the screened
self-energy correction for the $2p_{3/2}$ state of Li-like ions, which
concludes the calculation of all QED corrections of order $1/Z$ for the
$2p_{3/2}$-$2s$ transition. The remaining QED effect that involves two
virtual photons is the one-electron two-loop QED correction. Its calculation
to all order to $\Za$ is presently performed for the $1s$ state only
\cite{yerokhin:03:prl,yerokhin:03:epjd}. In this  work, we present an
estimation for this correction based on known terms of the $\Za$ expansion
and on the full result for the $1s$ state.

The paper is organized as follows. In Sec.~\ref{sec:formulas} we describe
general formulas for the screened self-energy correction for a Li-like ion. A
brief discussion of the renormalization of the formal expressions is given.
In Sec.~\ref{sec:results} we present the results of our numerical evaluation.
Various theoretical contributions to the $2p_{3/2}$-$2s$ transition energy
are collected in Sec.~\ref{sec:total}. Theoretical predictions obtained for
the total transition energy are compared with results by other authors and
experimental data. The relativistic units ($\hbar=c=m=1$) are used throughout
the paper.

\section{Basic formulas}
\label{sec:formulas}

\begin{figure}
\centering
\includegraphics[clip=true,width=0.9\textwidth]{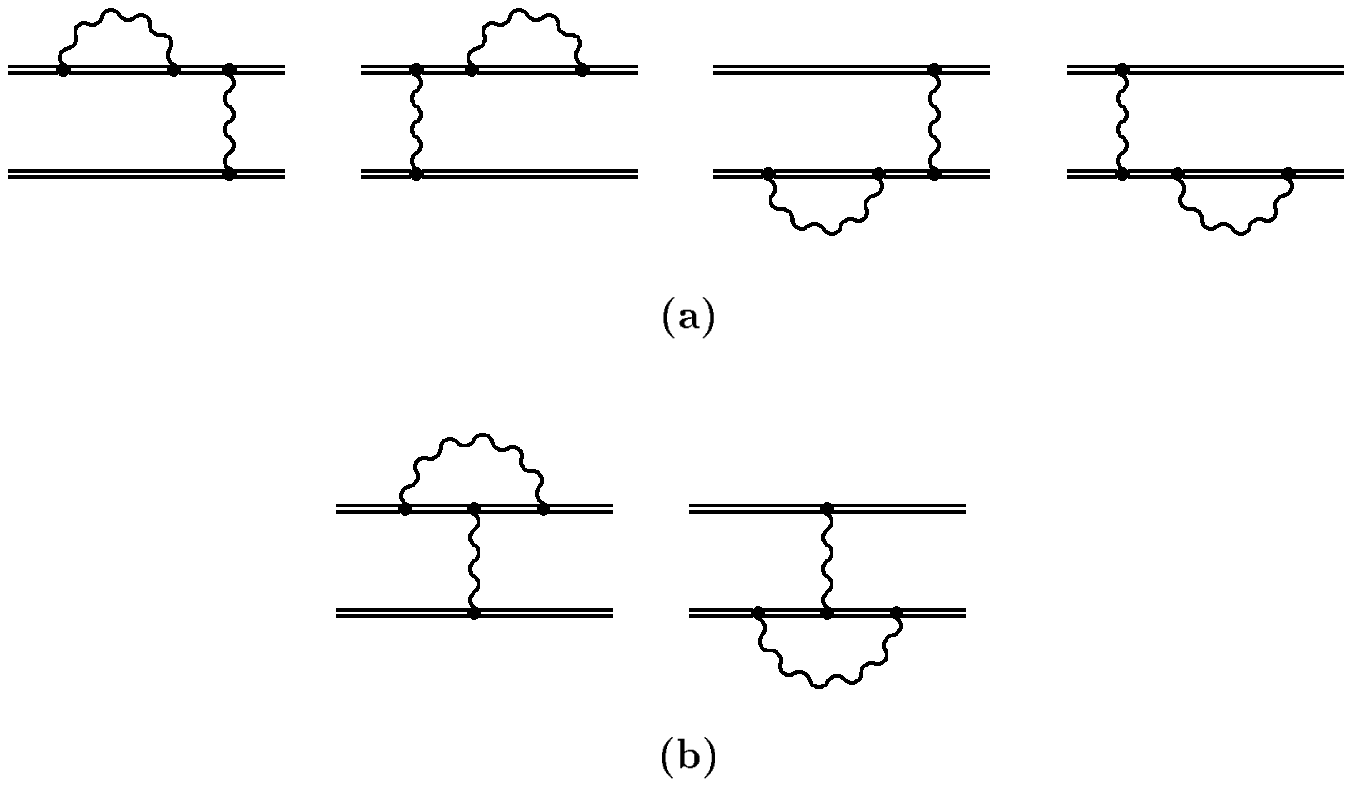}
\caption{Feynman diagrams representing the screened self-energy correction.
Double line indicates that the electron propagates in the field of the
nucleus. \label{fig:scrse} }
\end{figure}

The screened self-energy correction is graphically represented by Feynman
diagrams shown in Fig.~\ref{fig:scrse}. The detailed derivation of formal
expressions for this correction can be found in
Ref.~\cite{yerokhin:99:sescr}, where it was obtained by the two-time Green
function method developed by Shabaev (see Ref.~\cite{shabaev:02:rep} for the
description of the method). Here, we present only the final expressions for
the screened self-energy correction due to the interaction of the valence
electron with the $(1s)^2$ shell.

The contribution of diagrams in Fig.~\ref{fig:scrse}(a) is conveniently
divided into the {\em reducible} and {\em irreducible} parts. The reducible
part is a contribution in which the energy of the total intermediate state of
the system coincides with the initial (final) energy of the system, and the
irreducible part is the remainder. The irreducible part can be expressed in
terms of non-diagonal matrix elements of the one-loop self-energy operator
$\Sigma$,
\begin{equation} \label{1}
\Delta E_{\rm ir} = 2\sum_{\mu_c}\,\left[ \lbr c|\, \Sigma(\vare_c)\,|\delta
c\rbr
        + \lbr v|\,\Sigma(\vare_v)\,|\delta v\rbr \right] \, ,
\end{equation}
where $c$ and $v$ label the core and the valence electron, respectively;
$\mu_c$ is the momentum projection of the core electron, the factor of 2
accounts for two equivalent diagrams, the self-energy operator is defined by
\beqn   \label{sefunction}
\Sigma(\vare,\bfx_1,\bfx_2) &=& 2i\alpha \int^{\infty}_{-\infty} d\omega\,
    D^{\mu\nu}(\omega,\bfx_{12})\,
      \alpha_{\mu}\,
         G(\vare-\omega,\bfx_1,\bfx_2)\, \alpha_{\nu} \,,
\eeqn
$G$ is the Dirac-Coulomb Green function $G(\vare) = [\vare-H(1-i0)]^{-1}$,
$H$ is the Dirac-Coulomb Hamiltonian, $D^{\mu\nu}$ is the photon propagator,
${\alpha}^{\mu} = (1, \balpha)$ are the Dirac matrices, and $\bfx_{12} =
\bfx_1-\bfx_2$. The modified wave functions $|\delta c\rbr$ and $|\delta
v\rbr$ in Eq.~(\ref{1}) are the first-order perturbations of the initial wave
functions  $| c\rbr$ and $| v\rbr$ due to the electron-electron interaction:
\begin{eqnarray} \label{2}
|\delta c\rbr &=& \sum_n^{\vare_n \ne \vare_c} \frac{|n\rbr}{\vare_c-\vare_n}
        \sum_P (-1)^P \lbr Pc Pv |\,I(\Delta)\, |nv\rbr \, , \\
    \label{3}
|\delta v\rbr &=& \sum_n^{\vare_n \ne \vare_v} \frac{|n\rbr}{\vare_v-\vare_n}
        \sum_P (-1)^P \lbr Pc Pv |\,I(\Delta)\, |cn\rbr \, ,
\end{eqnarray}
where $I$ is the operator is the electron-electron interaction,
\beq  \label{Iomega}
I(\omega) = e^2 \alpha_{\mu}\alpha_{\nu} D^{\mu\nu}(\omega)\,,
\eeq
$P$ is the permutation operator [$(PcPv) = (cv)$, $(vc)$], and $ \Delta =
\vare_{Pc}-\vare_c$.

The reducible part is given by
\begin{eqnarray} \label{4}
\Delta E_{\rm red} &=& \sum_{\mu_c} \Bigl\{
 \sum_P (-1)^P \lbr PcPv|\,I(\Delta )\,|cv\rbr\,
 \left[\lbr c|\,{\Sigma}^{\prime}(\vare_c)\,|c\rbr +
       \lbr v|\,{\Sigma}^{\prime}(\vare_v)\,|v\rbr  \right]
\nonumber\\ &&
 + \lbr vc|\,I^{\prime}(\Delta )\,|cv\rbr
   \left[ \lbr c|\,{\Sigma}(\vare_c)\,|c\rbr
   - \lbr v|\,{\Sigma}(\vare_v)\,|v\rbr
   \right]\Bigr\}
       \, ,
\end{eqnarray}
where $I^{\prime}(\Delta ) =  \left. d /(d\omega) I(\omega) \right|_{\omega =
\Delta}$, $\Sigma^{\prime}(\vare_i) =  \left. d /(d\vare) \Sigma(\vare)
\right|_{\vare = \vare_i}$.

The contribution of the diagrams in Fig.~\ref{fig:scrse}(b) is referred to as
the {\em vertex part} and is given by
\begin{eqnarray} \label{5}
\Delta E_{\rm ver} &=& \sum_{\mu_c}\,\sum_P (-1)^P \sum_{n_1 n_2}
    \frac{i}{2\pi}
        \int^{\infty}_{-\infty} d\omega\, \left[
\frac{\lbr n_1 Pv|I(\Delta )|n_2 v\rbr
                \lbr Pc n_2|I(\omega )|n_1 c\rbr}
        {(\vare_{Pc} -\omega -u\vare_{n_1})
         (\vare_c -\omega -u\vare_{n_2})    } \right.
         \nonumber \\
&& + \left.
    \frac{ \lbr Pc n_1|I(\Delta )|cn_2\rbr
             \lbr Pvn_2| I(\omega )|n_1 v\rbr }
       {(\vare_{Pv} -\omega -u\vare_{n_1})
        (\vare_v -\omega -u\vare_{n_2})}
\right]\,,
\end{eqnarray}
where $u = 1-i0$ ensures the correct position of poles of the electron
propagators with respect to the integration contour.

The total self-energy correction to the interaction of the valence electron
with the $(1s)^2$ shell is given by the sum of the irreducible, reducible,
and vertex parts:
\beq
\Delta E_{\rm scr.\,se.} = \Delta E_{\rm ir}+ \Delta E_{\rm red}+ \Delta
E_{\rm ver}\,.
\eeq

The formulas presented so far are only the formal expressions and suffer from
ultraviolet (UV) and infrared (IR) divergences that should be explicitly
eliminated before the numerical calculation can be started. We note that the
irreducible part given by Eq.~(\ref{1}) is expressed in terms of non-diagonal
matrix elements of the one-loop self-energy operator and, therefore, can be
calculated by a straightforward generalization of a scheme developed for the
first-order self-energy correction. The method used for this in the present
investigation is based on an expansion of the bound-electron propagator in
terms of the interaction with the nuclear Coulomb field \cite{snyderman:91};
a detailed description of the numerical procedure is given in
Ref.~\cite{yerokhin:99:pra}. The second term in the reducible part (\ref{4})
consists of the first-order self-energy corrections multiplied by a simple
expression; the corresponding numerical evaluation is also reduced to a
first-order calculation.

The calculation of the remaining part of Eq.~(\ref{4}) and the vertex part
(\ref{5}) is more difficult. These two contributions should be considered
together since they contain UV and IR divergences that cancel each other in
the sum. In order to separate UV divergences in a covariant way, we separate
from the reducible and vertex parts the contribution of free-electron
propagators. This contribution is treated in momentum space within the
dimensional regularization and the divergences are separated by using the
standard technique of free-particle QED. The remainder does not contain any
UV divergences and is treated in coordinate space. IR divergences still
present in the remainder are separated, regularized by introducing a finite
photon mass, and canceled analytically. The general procedure for handling
divergent terms and the proof of their cancellation is described in our
previous investigation \cite{yerokhin:99:pra}.

\section{Numerical calculation and results}
\label{sec:results}

The calculation of the screened self-energy correction for the
$(1s)^22p_{3/2}$ state resembles that for other $n=2$ states described in our
previous work \cite{yerokhin:99:sescr}. A higher value of the total momentum
of the valence electron ($j_v = 3/2$) in the present case makes final
expressions more lengthy and their numerical evaluation more time consuming.
Significant complications appear when performing angular integrations in
momentum space for the vertex part with free electron propagators. For this
purpose, we developed a generalization of the angular-integration procedure
described in Ref.~\cite{yerokhin:99:sescr} to arbitrary states, using our
experience in calculating similar angular integrals for the one-electron
two-loop self-energy diagrams \cite{yerokhin:03:epjd}.

The actual calculation was carried out in the Feynman gauge and taking into
account the finite size of the nucleus. The homogeneously-charged
spherical-shell model was employed for the nuclear-charge distribution. Our
numerical results for the screened self-energy correction due to the
interaction of the valence electron with the $(1s)^2$ shell for the $2s$ and
$2p_{3/2}$ states of Li-like ions are presented in Table~\ref{tab:sescr} in
terms of the dimensionless function $F(\aZ)$ defined as
\beq    \label{FaZ}
\Delta E = m\,\alpha^2 (\aZ)^3F(\aZ)\,.
\eeq
The results  listed for the $2s$ state are very close to those obtained in
our previous investigation \cite{yerokhin:99:sescr}. In the present work, we
slightly improve the numerical accuracy and extend our calculation to $Z =
12$, 14, and 16. In the table, we compare our results with those by
Sapirstein and Cheng \cite{sapirstein:01:lamb} obtained for one specific case
$Z=83$. We observe a certain deviation of their numerical values from our
results. A similar disagreement is present also for the $2p_{1/2}$ state
\cite{yerokhin:99:sescr}. This discrepancy is not resolved at present. We
note, however, that our results for $n=2$ states of He-like ions
\cite{artemyev:04} (which are strongly related to the correction considered
here) agree well with the known terms of the $\Za$ expansion and that our
results for the ground state of He-like ions \cite{yerokhin:97:pla} are in
excellent agreement with an independent calculation by Sunnergren
\cite{sunnergren:98:phd}.

%
%
\begin{table}
\caption{Screened self-energy correction due to the interaction of the
valence electron with the $(1s)^2$ shell for the $2s$ and $2p_{3/2}$ states
of Li-like ions, in units of $F(\Za)$. $\lbr r^2\rbr^{1/2}$ is the
root-mean-square radius expressed in Fermi. \label{tab:sescr}}
\begin{tabular}{clr@{}lr@{}l} \hline
$Z$ &  $\lbr r^2\rbr^{1/2}$&
                            \multicolumn{2}{c}{$2s$}  &
                                   \multicolumn{2}{c}{$2p_{3/2}$}
\\ \hline
12   & 3.057 &    $-$0.&6966(7)   &   $-$0.&2104(10)  \\
14   & 3.123 &    $-$0.&6491(6)   &   $-$0.&1986(9)   \\
16   & 3.363 &    $-$0.&6093(5)   &   $-$0.&1884(6)   \\
18   & 3.427 &    $-$0.&5755(5)   &   $-$0.&1797(3)   \\
20   & 3.478 &    $-$0.&5466(4)   &   $-$0.&1723(3)   \\
30   & 3.928 &    $-$0.&4492(2)   &   $-$0.&1480(2)   \\
40   & 4.270 &    $-$0.&3968(2)   &   $-$0.&1353(2)   \\
50   & 4.655 &    $-$0.&3693(2)   &   $-$0.&1288(2)   \\
60   & 4.914 &    $-$0.&3590(2)   &   $-$0.&1261(2)   \\
70   & 5.317 &    $-$0.&3628(1)   &   $-$0.&1259(1)   \\
80   & 5.467 &    $-$0.&38103(5)  &   $-$0.&1276(1)   \\
83   & 5.533 &    $-$0.&38956(3)  &   $-$0.&1284(1)   \\
     &       &    $-$0.&3908$^a$  &   $-$0.&1350$^a$  \\
90   & 5.802 &    $-$0.&41585(2)  &   $-$0.&1306(1)   \\
92   & 5.860 &    $-$0.&42526(2)  &   $-$0.&1314(1)   \\
100  & 5.886 &    $-$0.&47372(2)  &   $-$0.&1347(1)   \\
\hline
\end{tabular}

\vspace*{0.2cm}
$^a$ Ref.~\cite{sapirstein:01:lamb}.
\end{table}

\section{2p$_{3/2}$-2s transition energy in Li-like ions}
\label{sec:total}

In this section we collect all presently available theoretical contributions
to the $2p_{3/2}$-$2s$ transition energy in Li-like bismuth, thorium, and
uranium. Individual corrections for these ions are presented in
Table~\ref{tab:total}. The {\em Dirac values} including the
finite-nuclear-size effect were obtained by solving the Dirac equation and
employing the two-parameter Fermi model for the nuclear-charge distribution.
Parameters of the Fermi model were expressed in terms of the root-mean-square
(rms) radii, which numerical values are listed in the table. The uncertainty
of the nuclear-size effect was evaluated by averaging two errors obtained by
varying the rms radius within the error bars given in the table and by
varying the model of the nuclear-charge distribution (the Fermi and the
homogeneously-charged-sphere models were employed). The {\em one-photon
exchange} correction was evaluated utilizing the Fermi model for the
nuclear-charge distribution. The one-loop {\em self-energy} correction was
taken from a tabulation in Ref.~\cite{beier:98:pra} for the $2s$ state and
from Ref.~\cite{mohr:92:b} for $2p_{3/2}$ state. The Uehling part of the
one-loop {\em vacuum-polarization} correction was calculated in this work for
the Fermi nuclear model. The Wichmann-Kroll part of this correction was taken
from a tabulation in Ref.~\cite{beier:97:jpb}. The {\em two-photon exchange}
correction was evaluated within framework of QED in our previous
investigation \cite{artemyev:03}. Numerical values for the {\em screened
self-energy} correction are taken from Table~\ref{tab:sescr}. The {\em
screened vacuum-polarization} correction was calculated in
Ref.~\cite{artemyev:99}.

Rigorous calculation of the {\em two-loop} QED correction is a challenging
problem, which is presently accomplished for the $1s$ state and for $Z \ge
40$ only \cite{yerokhin:03:prl,yerokhin:03:epjd}. For excited states, one has
to rely on the $\Za$ expansion, which reads
\beqn
\Delta E_{\rm 2loop} &=& m\,\frac{\alpha^2}{\pi^2}
 \frac{(\Za)^4}{n^3}
    \left\{ B_{40}+ (\Za) B_{50}+ (\Za)^2 \left[L^3B_{63}+
    \right.\right. \nonumber \\ && \left.\left.
    L^2B_{62}
             +L\,B_{61}+G^{\rm h.o.}(\Za)\right]  \right\}\,,
\eeqn
where $L = \ln [(\Za)^{-2}]$, $G^{\rm h.o.}(\Za) = B_{60}+ (\Za)(\cdots)$ is
the higher-order remainder. For $ns$ states, results for all coefficients up
to $B_{60}$ are available, whereas for $np$ states calculations were
performed for the coefficients up to $B_{62}$ only. (Details can be found in
a review \cite{mohr:00:rmp}, references therein, and more recent studies
\cite{pachucki:01:pra,pachucki:03:prl}.) Great care should be taken employing
the $\Za$ expansion for the evaluation of the total two-loop correction for
middle- and high-$Z$ ions, due to a very slow convergence of this expansion.
In order to estimate the two-loop QED correction for the $2s$ state, we
separate the $1s$ higher-order remainder $G^{\rm h.o.}(\Za)$ from the
numerical data of Ref.~\cite{yerokhin:03:epjd} and use it as an estimation of
the corresponding contribution for the $2s$ state, with an uncertainty of
50\%. For $p$ states, no analytical calculations for $B_{61}$ coefficient
exist up to now. We thus separate from the $1s$ numerical results of
Ref.~\cite{yerokhin:03:epjd} the function
\beq
 \widetilde{G}^{\rm h.o.}(\Za) = L\,B_{61}+
        G^{\rm h.o.}(\Za)\,,
\eeq
divide it by a factor of 8, and take the result as an uncertainty for the
higher-order contribution for $p$ states.

The relativistic {\em recoil} correction was evaluated to all orders in
$Z\alpha$ in Refs.~\cite{artemyev:95:pra,artemyev:95:jpb}. The {three-photon
exchange} correction was calculated for $Z=83$ in
Ref.~\cite{sapirstein:01:lamb} by utilizing many-body perturbation theory.
For two other cases, $Z=90$ and 92, we use the result for $Z=83$ with a 100\%
uncertainty. For $Z=83$, a 50\% uncertainty is assumed, which corresponds to
neglected QED effects. Finally, the {\em nuclear-polarization} correction was
calculated in Refs.~\cite{plunien:95,nefiodov:96}.

The total theoretical values for the transition energy in
Table~\ref{tab:total} are compared with the experimental data
\cite{beiersdorfer:98,beiersdorfer:95,beiersdorfer:95:nimb} and with the
previous theoretical evaluations
\cite{indelicato:90,kim:91,blundell:93:a,chen:95,sapirstein:01:lamb}. We note
that in all previous calculation except the one of
Ref.~\cite{sapirstein:01:lamb}, two-electron QED effects (effects of the
``screening" of QED corrections) were accounted for only approximately or
partly. A treatment which is closest to the approach presented in this work
is that by Sapirstein and Cheng \cite{sapirstein:01:lamb}, where all
two-electron QED corrections were evaluated for $Z=83$. Difference between
our total result and that of Ref.~\cite{sapirstein:01:lamb} is mainly due to
the estimate of the two-loop QED correction that is included in the present
compilation but was not accounted for in Ref.~\cite{sapirstein:01:lamb}.

Summarizing, we have presented an evaluation of the screened self-energy
correction to the $2p_{3/2}$-$2s$ transition energy of Li-like ions with
$Z\ge12$. This concludes our calculation of all two-electron QED corrections
for this transition and considerably improves the corresponding theoretical
predictions. It is demonstrated that the largest theoretical uncertainty for
high-$Z$ ions stems now from the two-loop QED correction, which calculation
to all orders in $\Za$ is needed in order to approach the experimental
accuracy.

%
%
\begin{table}
\caption{
Individual contributions to the $2p_{3/2}$-$2s$ transition energy
in Li-like ions, in eV.
\label{tab:total}
}
\begin{tabular}{lr@{}lr@{}lr@{}l}
\hline \hline
         & \multicolumn{2}{c}{$Z=83$}
                              & \multicolumn{2}{c}{$Z=90$}
                                               & \multicolumn{2}{c}{$Z=92$} \\
$\lbr r^2\rbr^{1/2}$ [Fm]
            &   \multicolumn{2}{c}{5.533(20)}
                              &  \multicolumn{2}{c}{5.802(4)}
                                                  &  \multicolumn{2}{c}{5.860(2)}     \\
\hline
Dirac value (extended nucleus) &   2792.&164(79)    &    4076.&681(65)    &     4527.&884(76)  \\
One-photon exchange  &     23.&821        &     -14.&459        &      -28.&411      \\
Self-energy          &    -35.&903        &     -51.&120        &      -56.&516      \\
Vacuum-polarization  &      8.&421        &      13.&561        &       15.&539      \\
Two-photon exchange  &     -1.&605        &      -0.&956        &       -0.&728      \\
Screened self-energy &      1.&579        &       2.&199        &        2.&420      \\
Screened vacuum polarization   &     -0.&431        &      -0.&680        &       -0.&776      \\
Two-loop QED         &      0.&21(19)     &       0.&30(23)     &        0.&33(24)   \\
Recoil               &     -0.&066        &      -0.&086        &       -0.&095      \\
Three-photon exchange&     -0.&024(12)    &      -0.&02(2)      &       -0.&02(2)    \\
Nuclear polarization &      0.&005(5)     &       0.&02(2)      &        0.&03(3)    \\
\hline
Total theory         &   2788.&17(21)     &    4025.&44(24)     &     4459.&66(25)   \\
Experiment           &   2788.&14(4)$^a$  &    4025.&23(14)$^b$ &     4459.&37(21)$^c$ \\
{\em Other calculations:} \\
Indelicato and Desclaux, 1990 \cite{indelicato:90}      &   2788.&2          &    4026.&3          &     4459.&9        \\
Kim \etal, 1991 \cite{kim:91}                           &   2787.&84         &    4024.&96         &     4459.&10       \\
Blundell, 1993 \cite{blundell:93:a}                     &        &           &    4025.&10         &          &         \\
Chen \etal, 1995 \cite{chen:95}                         &        &           &    4024.&98$^d$     &     4459.&13$^d$    \\
Sapirstein and Cheng, 2001 \cite{sapirstein:01:lamb}    &   2787.&96         &         &           &          &         \\
\hline \hline
\end{tabular}

\vspace*{0.2cm}
$\ ^a$ Ref.~\cite{beiersdorfer:98},
$\ ^b$ Ref.~\cite{beiersdorfer:95},
$\ ^c$ Ref.~\cite{beiersdorfer:95:nimb},\\
$\ ^d$ corrected for the right value of the nuclear-polarization correction
\cite{plunien:95,nefiodov:96}.
\end{table}

%
\section*{Acknowledgements}

This work was supported in part by RFBR (Grant No. 04-02-17574) and by the
Russian Ministry of Education (Grant No. E02-3.1-49). The work of V.M.S. was
supported by the Alexander von Humboldt Stiftung. V.A.Y. acknowledges the
support of the foundation "Dynasty". G.P. and G.S. acknowledge financial
support by the BMBF, DFG, and GSI.


\end{document}